\documentclass[pra,aps,showpacs,preprint,nofootinbib]{revtex4}
\usepackage{graphicx}
\usepackage{amsmath}

\begin{document}
\title{\bf The lower bound to the concurrence for four-qubit W state under noisy channels}
\author{Pakhshan Espoukeh}
\author{Pouria Pedram}
\email[Electronic address: ]{p.pedram@srbiau.ac.ir}
\affiliation{Department of Physics, Science and Research Branch,
Islamic Azad University, Tehran, Iran}
\date{\today}

\begin{abstract}
We study the dynamics of four-qubit W state under various noisy
environments by solving analytically the master equation in the
Lindblad form in which the Lindblad operators correspond to the
Pauli matrices and describe the decoherence of states. Also, we
investigate the dynamics of the entanglement using the lower bound
to the concurrence. It is found that while the entanglement
decreases monotonically for Pauli-Z noise, it decays suddenly for
other three noises. Moreover, by studying the time evolution of
entanglement of various maximally entangled four-qubit states, we
indicate that the four-qubit W state is more robust under same-axis
Pauli channels. Furthermore, three-qubit W state preserves more
entanglement with respect to the four-qubit W state, except for the
Pauli-Z noise.
\end{abstract}

 \maketitle

\section{Introduction}
The theory of open systems has an important role in quantum
information science \cite{breuer02}. The interaction between the
system and the environment which leads to the relaxation and
decoherence is the origin of the classical states. Therefore,
studying open quantum systems can help us to protect the quantum
states using proper methods
\cite{Misra77,vitali99,viola05,khodj07,kofman04,wu02}. The
application of quantum dynamics of open systems ranges from quantum
cosmology to quantum optics and to quantum information
\cite{2,3,4,7,8,9,10,11}. For instance, the detailed coherent
control over components of quantum systems is one of the greatest
challenges in quantum information processing. Among these efforts,
the control of entanglement against the dissipative effects of the
environment is an important issue.

The dynamics of open quantum systems is usually described by the
quantum Markov process as a semigroup of completely positive
dynamical maps and the corresponding quantum master equation in the
Lindblad form \cite{lind76,gorini76}. However, there are many
quantum systems which display non-Markovian behavior in which
correlations can give rise to memory effects due to the environment \cite{13,14,15}.
The Lindblad equation as the most general Markov evolution equation
for a density operator is given by \cite{lind76}
\begin{equation}\label{Lindblad}
\frac{\partial \rho}{\partial t} = -\frac{i}{\hbar} [H_S, \rho] +
\frac{1}{2} \left( [L,\rho L^{\dagger}]+ [L \rho, L^{\dagger}]
 \right),
\end{equation}
where the Lindblad operator $L$ represents the influence of the
environment.

Since the entanglement is known as a resource for performance
enhancements in quantum information processing, understanding and
control of the entanglement in open systems have many applications
in quantum computation and quantum information. As the first attempt
to study the dynamics of entanglement in open systems, the evolution
of entanglement has been studied using the state changes rather than
computing the entanglement of an arbitrary state by an entanglement
measure \cite{yi99}. The first real entanglement dynamics in open
systems investigated the dynamics of entanglement for a pair
initially entangled harmonic oscillators under the action of local
environments \cite{raj01}. Also, the evolution of the average
entanglement of formation of random bipartite states has been
analyzed in Ref.~\cite{zyc01}. In these works, the authors showed
that the entanglement vanishes at finite times and coherence decayed
asymptotically at long times.

The first study of dynamics of multipartite systems explored
multiqubit systems under the influence of single-qubit
depolarization and examined the robustness of three- and four-qubit
GHZ, W, and Dicke states and found that GHZ states are more robust
than other generic states \cite{simon02}. In Ref. \cite{siomau10},
in addition to depolarizing noise, the effects of Pauli noises are
studied on three-qubit GHZ and W states and discovered that
three-qubit GHZ state preserves more entanglement than three-qubit W
state. The dynamical evolution of $N$-qubit GHZ and W states for $2
\leq N\leq 7$ coupling to the independent dephasing and thermal
baths at zero temperature and infinite temperature has been compared
numerically using multipartite concurrence as a quantifier of the
entanglement. It is observed that the decay rate increases with $N$
for the GHZ state. For the W state this phenomenon happens only for
infinite temperature environment. Also, it is size-independent for
dephasing and zero-temperature thermal reservoirs \cite{carv04}. In
Ref.~\cite{SP1}, the robustness of $N$-qubit GHZ state as resources
for teleportation against Pauli channels and dephasing channel has
been studied. It is found that three-qubit GHZ state is more robust
than $N$-qubit ($N>3$) GHZ states under most noisy channels.  It is
also shown that three-qubit W state is more robust than three-qubit
GHZ state for small noisy parameter while the GHZ state becomes more
robust when the noisy parameter is large \cite{16}. Moreover, for W
state-like superpositions against dephasing and amplitude damping
channels, using three measures of entanglement, it is shown that the
effects of decoherence on the fidelity is better described by the
Meyer-Wallach global entanglement measure \cite{chaves10}.

In this paper, we study the changes of entanglement for the initial
four-qubit standard W state which will be mixed due to transmission
through noisy channels. For this purpose, we prepare initially a
pure W state then by solving analytically Eq.~(\ref{Lindblad}) with
Pauli matrices as the Lindblad operators we obtain dynamical
evolution of system under the influence of Pauli channels as well as
isotropic (depolarizing) channel. Also, to compute the entanglement,
we employ a lower bound for the multi-qubit concurrence proposed by
Li \emph{et al.} \cite{li09}. We note that the W state with the form
\begin{equation}\label{wstate}
|W_N\rangle=\frac{1}{\sqrt{N}}\left(|00\ldots01\rangle+|00\ldots10\rangle+\ldots+|01\ldots00\rangle+|10\ldots00\rangle
\right),
\end{equation}
is an example of multipartite entangled states in which $N$ denotes
the number of qubits in the state. Next, in order to compare the
robustness\footnote{The notion of ``robustness'' here and throughout
the paper is the persistence of the lower bound of entanglement in
noisy environments.} of $W_4$ state with maximally entangled
four-partite states, we examine the behavior of entanglement for
three four-qubit maximally entangled states that are given by
\cite{bassi}
\begin{eqnarray}\label{1}
&&|\phi_1\rangle=\frac{1}{\sqrt{2}}\left(|0000\rangle+|1111\rangle
\right),\\
&&|\phi_2\rangle=\frac{1}{2}\left(|1111\rangle+|1100\rangle+|0010\rangle+|0001\rangle
\right),\\
&&|\phi_3\rangle=\frac{1}{\sqrt{6}}\left(\sqrt{2}|1111\rangle+|1000\rangle+|0100\rangle+|0010\rangle+|0001\rangle
\right).
\end{eqnarray}

The rest of paper is as follows: In the next section, we
analytically solve the Lindblad equation for the initial four-qubit
W state and study the evolution of state under some Markovian
noises. Also, we study the behavior of entanglement of states under
the Pauli noises. In Sec.~\ref{sec3}, we investigate the time
evolution of entanglement for maximally entangled four-qubit states.
In the last section, we present our conclusions.

\begin{figure}[t]
%
%
\begin{picture}(350,150)
\put(-15,80){\resizebox{20pt}{13pt}{$|\psi\rangle$}}
\put(10,60){\resizebox{15pt}{75pt}{\{}}
\put(325,60){\resizebox{15pt}{75pt}{\}}}
\put(350,80){\resizebox{20pt}{15pt}{$|\bar{\psi}\rangle$}}

\put(40,130){\line(1,0){10}}\put(305,130){\line(1,0){10}}
\put(40,100){\line(1,0){10}}\put(305,100){\line(1,0){10}}
\put(40,70){\line(1,0){10}}\put(305,70){\line(1,0){10}}
\put(40,40){\line(1,0){10}}\put(305,40){\line(1,0){10}}


\put(50,30){\framebox(255,20){Noise}}
\put(50,60){\framebox(255,20){Noise}}
\put(50,90){\framebox(255,20){Noise}}
\put(50,120){\framebox(255,20){Noise}}

\put(168,3){time}

\thicklines
\put(10,0){\vector(1,0){340}}

\end{picture}
\caption{\label{circuit} A schematic quantum circuit for the
transmission of a four-qubit state through a noisy environment.}
\end{figure}
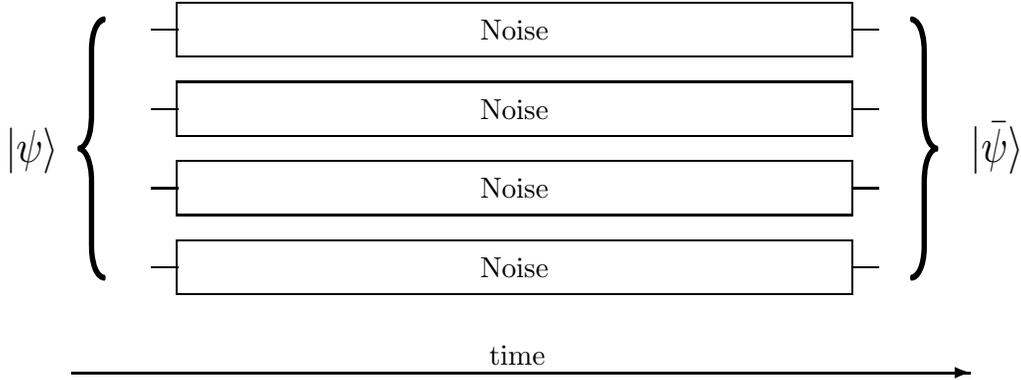

\section{Evolution of entanglement of $W_4$ state under noisy channels}\label{sec2}
In this section, we analytically solve the Lindblad equation
(\ref{Lindblad}) for the initial $W_4$ state in contact with some
Markovian noises and study the dynamics of the entanglement of the
states. It is remarkable to point that the Lindblad equation for a
four-particle state consists of 136 coupled differential equations
which in principle their solution is a difficult task. Here, using
the evolution of the density matrix at small times, we propose a
proper ansatz for the sought-after density matrix which leads to at
most 11 coupled equations that is much simpler to solve. Fig.
(\ref{circuit}) shows the quantum circuit that describes
schematically the effect of noise on the state.

\begin{figure}
\begin{center}
\includegraphics[width=10cm]{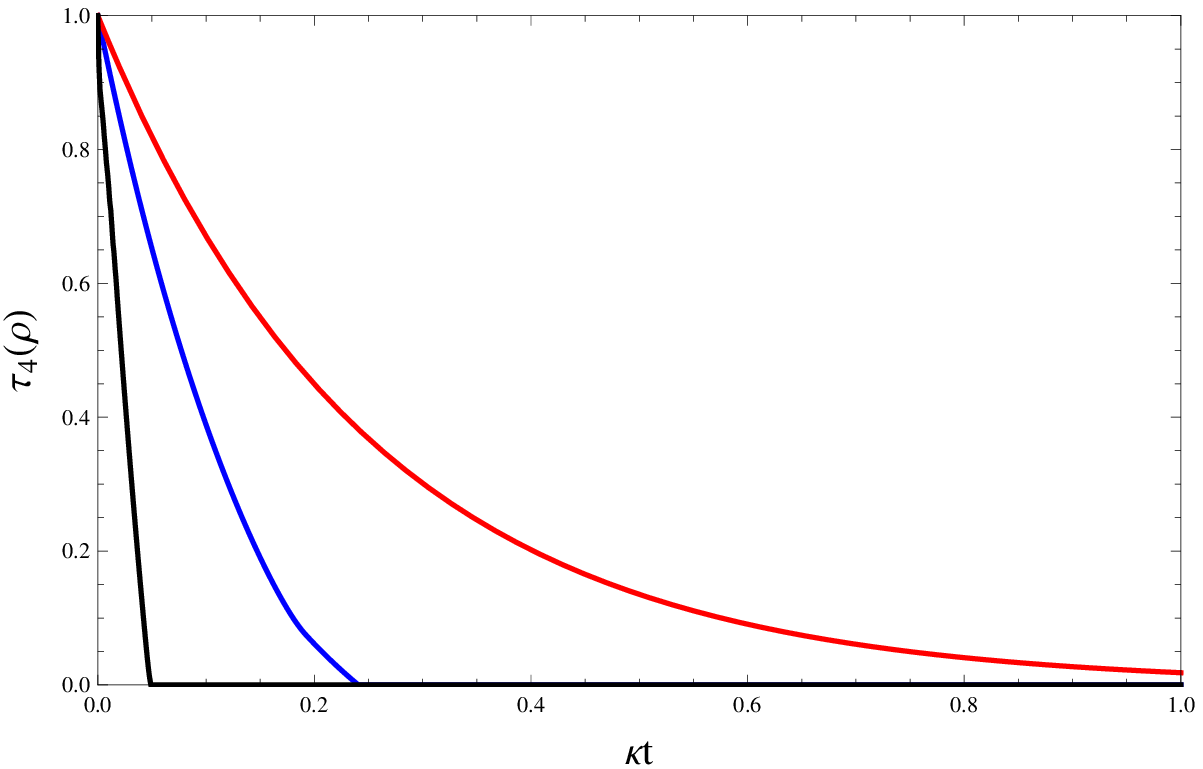}
\caption{\label{fig2} The lower bound for the four-qubit concurrence
for an initial W state transmitted through Pauli-X and Pauli-Y (blue
line), Pauli-Z (red line), and isotropic (black line) channels as
functions of $\kappa t$.}
\end{center}
\end{figure}

For the first case, consider Pauli-X (bit-flip) channel. When a
four-qubit W state transmits through this noise,
Eq.~(\ref{Lindblad}) leads to 16 diagonal and 120 off-diagonal
coupled linear differential equations. Thus, to simplify the
problem, we first consider the time evolution of the density matrix
$W_4$ state for infinitesimal time interval $\delta t$ as
\begin{equation}\label{del}
\rho(\delta t)= \rho(0) + \left[ \sum_{i=1}^4 \left(L_{i,x}
\rho(0) L_{i,x}^{\dagger} \right) - \frac{1}{2} \left\{
L_{i,x}^{\dagger} L_{i,x}, \rho(0) \right\} \right]\delta
t,
\end{equation}
in which $L_{i,x}=\sqrt{\kappa _{i,x}}\sigma_x^i$ and
\begin{eqnarray}\label{mat0}
\rho(0) = { \left(
\begin{smallmatrix}
0 & 0 & 0 & 0 & 0 & 0 & 0 & 0 & 0 & 0 & 0 & 0 & 0 & 0 & 0 & 0   \\
0 & 1 & 1 & 0 & 1 & 0 & 0 & 0 & 1 & 0 & 0 & 0 & 0 & 0 & 0 & 0   \\
0 & 1 & 1 & 0 & 1 & 0 & 0 & 0 & 1 & 0 & 0 & 0 & 0 & 0 & 0 & 0   \\
0 & 0 & 0 & 0 & 0 & 0 & 0 & 0 & 0 & 0 & 0 & 0 & 0 & 0 & 0 & 0   \\
0 & 1 & 1 & 0 & 1 & 0 & 0 & 0 & 1 & 0 & 0 & 0 & 0 & 0 & 0 & 0   \\
0 & 0 & 0 & 0 & 0 & 0 & 0 & 0 & 0 & 0 & 0 & 0 & 0 & 0 & 0 & 0   \\
0 & 0 & 0 & 0 & 0 & 0 & 0 & 0 & 0 & 0 & 0 & 0 & 0 & 0 & 0 & 0   \\
0 & 0 & 0 & 0 & 0 & 0 & 0 & 0 & 0 & 0 & 0 & 0 & 0 & 0 & 0 & 0   \\
0 & 1 & 1 & 0 & 1 & 0 & 0 & 0 & 1 & 0 & 0 & 0 & 0 & 0 & 0 & 0   \\
0 & 0 & 0 & 0 & 0 & 0 & 0 & 0 & 0 & 0 & 0 & 0 & 0 & 0 & 0 & 0   \\
0 & 0 & 0 & 0 & 0 & 0 & 0 & 0 & 0 & 0 & 0 & 0 & 0 & 0 & 0 & 0   \\
0 & 0 & 0 & 0 & 0 & 0 & 0 & 0 & 0 & 0 & 0 & 0 & 0 & 0 & 0 & 0   \\
0 & 0 & 0 & 0 & 0 & 0 & 0 & 0 & 0 & 0 & 0 & 0 & 0 & 0 & 0 & 0   \\
0 & 0 & 0 & 0 & 0 & 0 & 0 & 0 & 0 & 0 & 0 & 0 & 0 & 0 & 0 & 0   \\
0 & 0 & 0 & 0 & 0 & 0 & 0 & 0 & 0 & 0 & 0 & 0 & 0 & 0 & 0 & 0   \\
0 & 0 & 0 & 0 & 0 & 0 & 0 & 0 & 0 & 0 & 0 & 0 & 0 & 0 & 0 & 0
\end{smallmatrix} \right).}
\end{eqnarray}
Substituting $\rho(0)$ in Eq.~(\ref{del}) gives
\begin{eqnarray}
\rho_{W_4}^x(\delta t) =
\frac{1}{4}{\left(
\begin{smallmatrix}
4\kappa \delta t & 0 & 0 & 2\kappa \delta t & 0 & 2\kappa \delta t & 2\kappa \delta t & 0 & 0 & 2\kappa \delta t & 2\kappa \delta t & 0 & 2\kappa \delta t & 0 & 0 & 0 \\
0 & 1 - 4\kappa \delta t & 1 - 4\kappa \delta t & 0 & 1 - 4\kappa \delta t & 0 & 0 & 0 & 1 - 4\kappa \delta t & 0 & 0 & 0 & 0 & 0 & 0 & 0  \\
0 & 1 - 4\kappa \delta t & 1 - 4\kappa \delta t & 0 & 1 - 4\kappa \delta t & 0 & 0 & 0 & 1 - 4\kappa \delta t & 0 & 0 & 0 & 0 & 0 & 0 & 0  \\
2\kappa \delta t & 0 & 0 & 2\kappa \delta t & 0 & \kappa \delta t & \kappa \delta t & 0 & 0 & \kappa \delta t & \kappa \delta t & 0 & 0 & 0 & 0 & 0 \\
0 & 1 - 4\kappa \delta t & 1 - 4\kappa \delta t & 0 & 1 - 4\kappa \delta t & 0 & 0 & 0 & 1 - 4\kappa \delta t & 0 & 0 & 0 & 0 & 0 & 0 & 0  \\
2\kappa \delta t & 0 & 0 & \kappa \delta t & 0 & 2\kappa \delta t & \kappa \delta t & 0 & 0 & \kappa \delta t & 0 & 0 & \kappa \delta t & 0 & 0 & 0 \\
2\kappa \delta t & 0 & 0 & \kappa \delta t & 0 & \kappa \delta t & 2\kappa \delta t & 0 & 0 & 0 & \kappa \delta t & 0 & \kappa \delta t & 0 & 0 & 0 \\
0 & 0 & 0 & 0 & 0 & 0 & 0 & 0 & 0 & 0 & 0 & 0 & 0 & 0 & 0 & 0 \\
0 & 1 - 4\kappa \delta t & 1 - 4\kappa \delta t & 0 & 1 - 4\kappa \delta t & 0 & 0 & 0 & 1 - 4\kappa \delta t & 0 & 0 & 0 & 0 & 0 & 0 & 0  \\
2 \kappa \delta t & 0 & 0 & \kappa \delta t & 0 & \kappa \delta t  & 0 & 0 & 0 & 2\kappa \delta t & \kappa \delta t & 0 & \kappa \delta t & 0 & 0 & 0 \\
2 \kappa \delta t & 0 & 0 & \kappa \delta t & 0 & 0 & \kappa \delta t & 0 & 0 & \kappa \delta t & 2\kappa \delta t & 0 & \kappa \delta t & 0 & 0 & 0 \\
0 & 0 & 0 & 0 & 0 & 0 & 0 & 0 & 0 & 0 & 0 & 0 & 0 & 0 & 0 & 0 \\
2 \kappa \delta t & 0 & 0 & 0 & 0 & \kappa \delta t & \kappa \delta t & 0 & 0 & \kappa \delta t & \kappa \delta t & 0 & 2 \kappa \delta t & 0 & 0 & 0  \\
0 & 0 & 0 & 0 & 0 & 0 & 0 & 0 & 0 & 0 & 0 & 0 & 0 & 0 & 0 & 0 \\
0 & 0 & 0 & 0 & 0 & 0 & 0 & 0 & 0 & 0 & 0 & 0 & 0 & 0 & 0 & 0 \\
0 & 0 & 0 & 0 & 0 & 0 & 0 & 0 & 0 & 0 & 0 & 0 & 0 & 0 & 0 & 0
\end{smallmatrix}\right).}
\end{eqnarray}
Now, consider the following ansatz for all times which is consistent
with $\rho_{W_4}^x(\delta t) $:
\begin{eqnarray}\label{mat1}
\rho_{ {W_4}}^x (t)= { \left(
\begin{smallmatrix}
a & 0 & 0 & d & 0 & d & d & 0 & 0 & d & d & 0 & d & 0 & 0 & d   \\
0 & b & c & 0 & c & 0 & 0 & p & c & 0 & 0 & p & 0 & p & 0 & 0   \\
0 & c & b & 0 & c & 0 & 0 & p & c & 0 & 0 & p & 0 & 0 & p & 0   \\
d & 0 & 0 & e & 0 & m & m & 0 & 0 & m & m & 0 & 0 & 0 & 0 & q   \\
0 & c & c & 0 & b & 0 & 0 & p & c & 0 & 0 & 0 & 0 & p & p & 0   \\
d & 0 & 0 & m & 0 & e & m & 0 & 0 & m & 0 & 0 & m & 0 & 0 & q   \\
d & 0 & 0 & m & 0 & m & e & 0 & 0 & 0 & m & 0 & m & 0 & 0 & q   \\
0 & p & p & 0 & p & 0 & 0 & f & 0 & 0 & 0 & n & 0 & n & n & 0   \\
0 & c & c & 0 & c & 0 & 0 & 0 & b & 0 & 0 & p & 0 & p & p & 0   \\
d & 0 & 0 & m & 0 & m & 0 & 0 & 0 & e & m & 0 & m & 0 & 0 & q   \\
d & 0 & 0 & m & 0 & 0 & m & 0 & 0 & m & e & 0 & m & 0 & 0 & q   \\
0 & p & p & 0 & 0 & 0 & 0 & n & p & 0 & 0 & f & 0 & n & n & 0   \\
d & 0 & 0 & 0 & 0 & m & m & 0 & 0 & m & m & 0 & e & 0 & 0 & q   \\
0 & p & 0 & 0 & p & 0 & 0 & n & p & 0 & 0 & n & 0 & f & n & 0   \\
0 & 0 & p & 0 & p & 0 & 0 & n & p & 0 & 0 & n & 0 & n & f & 0   \\
0 & 0 & 0 & q & 0 & q & q & 0 & 0 & q & q & 0 & q & 0 & 0 & h
\end{smallmatrix} \right).}
\end{eqnarray}
Inserting this solution in the Lindblad equation results is a set of
11 coupled differential equations
\begin{eqnarray}\label{diff1}
\left\{
\begin{array}{l}
\dot{a}(t) = 4k\Big(b(t)-a(t)\Big),\\
\dot{b}(t) = k\Big(a(t)-4b(t)+3c(t)\Big),\\
\dot{c}(t)=2k\Big(d(t)+m(t)-2c(t)\Big),\\
\dot{d}(t)=2k\Big(c(t)+p(t)-2d(t)\Big),\\
\dot{e}(t)=2k\Big(b(t)+f(t)-2e(t)\Big),\\
\dot{f}(t) = k\Big(3e(t)+h(t)-4f(t)\Big),\\
\dot{h}(t) = 4k\Big(f(t)-h(t)\Big),\\
\dot{p}(t)=k\Big(d(t)+2m(t)-4p(t)+q(t)\Big),\\
\dot{m}(t)=k\Big(c(t)-4m(t)+n(t)+2p(t)\Big),\\
\dot{n}(t)=2k\Big(m(t)+q(t)-2n(t)\Big),\\
\dot{q}(t)=2k\Big(n(t)+p(t)-2q(t)\Big),
\end{array}\right.
\end{eqnarray}
subject to the initial conditions $b(0)=c(0)=1/4$ and
$a(0)=d(0)=e(0)=f(0)=h(0)=p(0)=m(0)=n(0)=q(0)=0$. The solutions read
\begin{eqnarray}
\label{solution1} \left\{
\begin{array}{l}
a(t) = 2d(t) =\frac{1}{16}\left( 1 + 2e^{-2 \kappa t} -  2e^{-6 \kappa t} - e^{-8 \kappa t}\right),\\
b(t) =\frac{1}{16}\left( 1+ e^{-2 \kappa t} +  e^{-6 \kappa t} + e^{-8 \kappa t}\right),\\
c(t) =\frac{1}{32}\left( 1 + 2e^{-2 \kappa t} + 2e^{-4 \kappa t} + 2 e^{-6 \kappa t} + e^{-8 \kappa t} \right),\\
e(t) = 2m(t) =\frac{1}{16}\left( 1 - e^{-8 \kappa t}\right),\\
f(t) =\frac{1}{16}\left( 1 - e^{-2 \kappa t} -  e^{-6 \kappa t} + e^{-8 \kappa t}\right),\\
h(t) = 2q(t) =\frac{1}{16}\left( 1 - 2e^{-2 \kappa t} +  2e^{-6 \kappa t} - e^{-8 \kappa t}\right),\\
n(t) =\frac{1}{32}\left( 1 - 2e^{-2 \kappa t} + 2e^{-4 \kappa t} - 2 e^{-6 \kappa t} + e^{-8 \kappa t}\right),\\
p(t) = \frac{1}{32}\left( 1 - 2e^{-4 \kappa t} + e^{-8 \kappa t}\right).
\end{array}\right.
\end{eqnarray}
Now, using $\rho_{ {W_4}}^x (t)$ we examine the evolution of the
entanglement for this state. As a quantifier of the entanglement, we
use the lower bound to multipartite concurrence which is introduced
by Li \emph{et al.} \cite{li09}
\begin{equation}\label{lowerbound}
C_N(\rho) \geq
\tau_N(\rho)\equiv\sqrt{\frac{1}{N}\sum_{n=1}^N\sum_{k=1}^K
(C_k^n)^2},
\end{equation}
that it is given in terms of $N$ bipartite concurrences $C^n$
correspond to the possible bipartite cuts of the multi-qubit system
in which one of qubits is separated from the other qubits. The
bipartite concurrence $C^n$ is defined by a sum of $K = 2^{N-2}
(2^{N-1} - 1)$ terms which are expressed as $C_k^n
=\mathrm{max}\{0,\lambda_k^1-\lambda_k^2-\lambda_k^3-\lambda_k^4\}$,
where $\lambda_k^m$, $m=1,...,4$, are the square roots of four
nonvanishing eigenvalues in decreasing order of the matrix $\rho
\tilde{\rho}_k^n$. Here, $\tilde{\rho}_k^n=S_k^n\rho^*S_k^n$ and
$S_k^n$ are defined by the tensor product of the generators of SO(2)
and SO($2^{N-1}$). For this case, the lower bound can be obtained
numerically which is plotted in Fig.~\ref{fig2} (blue line).

For the next case, consider the Pauli-Y (bit-phase-flip) noise.  Calculations in this
case show that results are similar to the previous case and
therefore both noises have the same effect on the four-qubit W
state, namely $\rho_{ {W_4}}^y (t)=\rho_{ {W_4}}^x (t)$.

Now, consider the case that $W_4$ state is coupled to the Pauli-Z (phase-flip)
channel. For this case, the infinitesimal time evolution of the
density matrix reads
\begin{eqnarray}
\rho_{W_4}^z(\delta t) =
\frac{1}{4}{\left(
\begin{smallmatrix}
0 & 0 & 0 & 0 & 0 & 0 & 0 & 0 & 0 & 0 & 0 & 0 & 0 & 0 & 0 & 0 \\
0 & 1 & 1 - 4\kappa \delta t & 0 & 1 - 4\kappa \delta t & 0 & 0 & 0 & 1 - 4\kappa \delta t & 0 & 0 & 0 & 0 & 0 & 0 & 0  \\
0 & 1 - 4\kappa \delta t & 1 & 0 & 1 - 4\kappa \delta t & 0 & 0 & 0 & 1 - 4\kappa \delta t & 0 & 0 & 0 & 0 & 0 & 0 & 0  \\
0 & 0 & 0 & 0 & 0 & 0 & 0 & 0 & 0 & 0 & 0 & 0 & 0 & 0 & 0 & 0 \\
0 & 1 - 4\kappa \delta t & 1 - 4\kappa \delta t & 0 & 1 & 0 & 0 & 0 & 1 - 4\kappa \delta t & 0 & 0 & 0 & 0 & 0 & 0 & 0  \\
0 & 0 & 0 & 0 & 0 & 0 & 0 & 0 & 0 & 0 & 0 & 0 & 0 & 0 & 0 & 0 \\
0 & 0 & 0 & 0 & 0 & 0 & 0 & 0 & 0 & 0 & 0 & 0 & 0 & 0 & 0 & 0 \\
0 & 0 & 0 & 0 & 0 & 0 & 0 & 0 & 0 & 0 & 0 & 0 & 0 & 0 & 0 & 0 \\
0 & 1 - 4\kappa \delta t & 1 - 4\kappa \delta t & 0 & 1 - 4\kappa \delta t & 0 & 0 & 0 & 1 & 0 & 0 & 0 & 0 & 0 & 0 & 0  \\
0 & 0 & 0 & 0 & 0 & 0 & 0 & 0 & 0 & 0 & 0 & 0 & 0 & 0 & 0 & 0 \\
0 & 0 & 0 & 0 & 0 & 0 & 0 & 0 & 0 & 0 & 0 & 0 & 0 & 0 & 0 & 0 \\
0 & 0 & 0 & 0 & 0 & 0 & 0 & 0 & 0 & 0 & 0 & 0 & 0 & 0 & 0 & 0 \\
0 & 0 & 0 & 0 & 0 & 0 & 0 & 0 & 0 & 0 & 0 & 0 & 0 & 0 & 0 & 0 \\
0 & 0 & 0 & 0 & 0 & 0 & 0 & 0 & 0 & 0 & 0 & 0 & 0 & 0 & 0 & 0 \\
0 & 0 & 0 & 0 & 0 & 0 & 0 & 0 & 0 & 0 & 0 & 0 & 0 & 0 & 0 & 0 \\
0 & 0 & 0 & 0 & 0 & 0 & 0 & 0 & 0 & 0 & 0 & 0 & 0 & 0 & 0 & 0
\end{smallmatrix}\right),}
\end{eqnarray}
so we take the ansatz
\begin{eqnarray}\label{mat2}
\rho_{{W_4}}^z (t)= { \left(
\begin{smallmatrix}
0 & 0 & 0 & 0 & 0 & 0 & 0 & 0 & 0 & 0 & 0 & 0 & 0 & 0 & 0 & 0   \\
0 & a & b & 0 & b & 0 & 0 & 0 & b & 0 & 0 & 0 & 0 & 0 & 0 & 0   \\
0 & b & a & 0 & b & 0 & 0 & 0 & b & 0 & 0 & 0 & 0 & 0 & 0 & 0   \\
0 & 0 & 0 & 0 & 0 & 0 & 0 & 0 & 0 & 0 & 0 & 0 & 0 & 0 & 0 & 0   \\
0 & b & b & 0 & a & 0 & 0 & 0 & b & 0 & 0 & 0 & 0 & 0 & 0 & 0   \\
0 & 0 & 0 & 0 & 0 & 0 & 0 & 0 & 0 & 0 & 0 & 0 & 0 & 0 & 0 & 0   \\
0 & 0 & 0 & 0 & 0 & 0 & 0 & 0 & 0 & 0 & 0 & 0 & 0 & 0 & 0 & 0   \\
0 & 0 & 0 & 0 & 0 & 0 & 0 & 0 & 0 & 0 & 0 & 0 & 0 & 0 & 0 & 0   \\
0 & b & b & 0 & b & 0 & 0 & 0 & a & 0 & 0 & 0 & 0 & 0 & 0 & 0   \\
0 & 0 & 0 & 0 & 0 & 0 & 0 & 0 & 0 & 0 & 0 & 0 & 0 & 0 & 0 & 0   \\
0 & 0 & 0 & 0 & 0 & 0 & 0 & 0 & 0 & 0 & 0 & 0 & 0 & 0 & 0 & 0   \\
0 & 0 & 0 & 0 & 0 & 0 & 0 & 0 & 0 & 0 & 0 & 0 & 0 & 0 & 0 & 0   \\
0 & 0 & 0 & 0 & 0 & 0 & 0 & 0 & 0 & 0 & 0 & 0 & 0 & 0 & 0 & 0   \\
0 & 0 & 0 & 0 & 0 & 0 & 0 & 0 & 0 & 0 & 0 & 0 & 0 & 0 & 0 & 0   \\
0 & 0 & 0 & 0 & 0 & 0 & 0 & 0 & 0 & 0 & 0 & 0 & 0 & 0 & 0 & 0   \\
0 & 0 & 0 & 0 & 0 & 0 & 0 & 0 & 0 & 0 & 0 & 0 & 0 & 0 & 0 & 0
\end{smallmatrix} \right).}
\end{eqnarray}
Inserting $\rho_{{W_4}}^z$ to the Lindblad equation (\ref{Lindblad})
gives rise in
\begin{eqnarray}\label{diff2}
\left\{
\begin{array}{l}
\dot{a}(t) = 0,\\
\dot{b}(t) =-4k\, b(t) ,
\end{array}\right.
\end{eqnarray}
subject to the initial conditions $a(0)=b(0)=1/4$. Then, we obtain
\begin{eqnarray}
\label{solution2} \left\{
\begin{array}{l}
a(t) =\frac{1}{4},\\
b(t) =\frac{1}{4} e^{-4 \kappa t}.
\end{array}\right.
\end{eqnarray}
The lower bound in this case is given by
\begin{equation}\label{entz}
\tau_4(\rho_{W}^z) = e^{-4 \kappa t}.
\end{equation}
This result is depicted in Fig.~\ref{fig2} as a red line.

To this end, we study the interaction of an initial four-qubit W
state with the depolarizing channel. Taking into account the effects
of all Pauli matrices as noisy operators in the Lindblad equation,
the infinitesimal time evolution of the density matrix becomes
\begin{eqnarray}
\rho_{W_4}^d(\delta t) =
\frac{1}{4}{\left(
\begin{smallmatrix}
2\kappa \delta t & 0 & 0 & 0 & 0 & 0 & 0 & 0 & 0 & 0 & 0 & 0 & 0 & 0 & 0 & 0 \\
0 & 1 - 8\kappa \delta t & 1 - 12\kappa \delta t & 0 & 1 - 12\kappa \delta t & 0 & 0 & 0 & 1 - 12\kappa \delta t & 0 & 0 & 0 & 0 & 0 & 0 & 0  \\
0 & 1 - 12\kappa \delta t & 1 - 8\kappa \delta t & 0 & 1 - 12\kappa \delta t & 0 & 0 & 0 & 1 - 12\kappa \delta t & 0 & 0 & 0 & 0 & 0 & 0 & 0  \\
0 & 0 & 0 & 4\kappa \delta t & 0 & 2\kappa \delta t & 2\kappa \delta t & 0 & 0 & 2\kappa \delta t & 2\kappa \delta t & 0 & 0 & 0 & 0 & 0 \\
0 & 1 - 12\kappa \delta t & 1 - 12\kappa \delta t & 0 & 1 - 8\kappa \delta t & 0 & 0 & 0 & 1 - 12\kappa \delta t & 0 & 0 & 0 & 0 & 0 & 0 & 0  \\
0 & 0 & 0 & 8\kappa \delta t & 0 & 4\kappa \delta t & 8\kappa \delta t & 0 & 0 & 8\kappa \delta t & 0 & 0 & 8\kappa \delta t & 0 & 0 & 0 \\
0 & 0 & 0 & 8\kappa \delta t & 0 & 8\kappa \delta t & 4\kappa \delta t & 0 & 0 & 0 & 8\kappa \delta t & 0 & 8\kappa \delta t & 0 & 0 & 0 \\
0 & 0 & 0 & 0 & 0 & 0 & 0 & 0 & 0 & 0 & 0 & 0 & 0 & 0 & 0 & 0 \\
0 & 1 - 12\kappa \delta t & 1 - 12\kappa \delta t & 0 & 1 - 12\kappa \delta t & 0 & 0 & 0 & 1 - 8\kappa \delta t & 0 & 0 & 0 & 0 & 0 & 0 & 0  \\
0 & 0 & 0 & 2\kappa \delta t & 0 & 2\kappa \delta t & 0 & 0 & 0 & 4\kappa \delta t & 2\kappa \delta t & 0 & 2\kappa \delta t & 0 & 0 & 0 \\
0 & 0 & 0 & 2\kappa \delta t & 0 & 0 & 2\kappa \delta t & 0 & 0 & 2\kappa \delta t & 4\kappa \delta t & 0 & 2\kappa \delta t & 0 & 0 & 0 \\
0 & 0 & 0 & 0 & 0 & 0 & 0 & 0 & 0 & 0 & 0 & 0 & 0 & 0 & 0 & 0 \\
0 & 0 & 0 & 0 & 0 & 2\kappa \delta t & 2\kappa \delta t & 0 & 0 & 2\kappa \delta t & 2\kappa \delta t & 0 & 4\kappa \delta t & 0 & 0 & 0 \\
0 & 0 & 0 & 0 & 0 & 0 & 0 & 0 & 0 & 0 & 0 & 0 & 0 & 0 & 0 & 0 \\
0 & 0 & 0 & 0 & 0 & 0 & 0 & 0 & 0 & 0 & 0 & 0 & 0 & 0 & 0 & 0 \\
0 & 0 & 0 & 0 & 0 & 0 & 0 & 0 & 0 & 0 & 0 & 0 & 0 & 0 & 0 & 0
\end{smallmatrix}\right).}
\end{eqnarray}
Therefore, we take the following ansatz:
\begin{eqnarray}\label{mat3}
\rho_{ {W_4}}^d = { \left(
\begin{smallmatrix}
g & 0 & 0 & 0 & 0 & 0 & 0 & 0 & 0 & 0 & 0 & 0 & 0 & 0 & 0 & 0   \\
0 & a & b & 0 & b & 0 & 0 & 0 & b & 0 & 0 & 0 & 0 & 0 & 0 & 0   \\
0 & b & a & 0 & b & 0 & 0 & 0 & b & 0 & 0 & 0 & 0 & 0 & 0 & 0   \\
0 & 0 & 0 & e & 0 & c & c & 0 & 0 & c & c & 0 & 0 & 0 & 0 & 0   \\
0 & b & b & 0 & a & 0 & 0 & 0 & b & 0 & 0 & 0 & 0 & 0 & 0 & 0   \\
0 & 0 & 0 & c & 0 & e & c & 0 & 0 & c & 0 & 0 & c & 0 & 0 & 0   \\
0 & 0 & 0 & c & 0 & c & e & 0 & 0 & 0 & c & 0 & c & 0 & 0 & 0   \\
0 & 0 & 0 & 0 & 0 & 0 & 0 & h & 0 & 0 & 0 & d & 0 & d & d & 0   \\
0 & b & b & 0 & b & 0 & 0 & 0 & a & 0 & 0 & 0 & c & 0 & 0 & 0   \\
0 & 0 & 0 & c & 0 & c & 0 & 0 & 0 & e & c & 0 & c & 0 & 0 & 0   \\
0 & 0 & 0 & c & 0 & 0 & c & 0 & 0 & c & e & 0 & 0 & 0 & 0 & 0   \\
0 & 0 & 0 & 0 & 0 & 0 & 0 & d & 0 & 0 & 0 & h & 0 & d & d & 0   \\
0 & 0 & 0 & 0 & 0 & c & c & 0 & 0 & c & c & 0 & e & 0 & 0 & 0   \\
0 & 0 & 0 & 0 & 0 & 0 & 0 & d & 0 & 0 & 0 & d & 0 & h & d & 0   \\
0 & 0 & 0 & 0 & 0 & 0 & 0 & d & 0 & 0 & 0 & d & 0 & d & h & 0   \\
0 & 0 & 0 & 0 & 0 & 0 & 0 & 0 & 0 & 0 & 0 & 0 & 0 & 0 & 0 & f
\end{smallmatrix} \right).}
\end{eqnarray}
Inserting this density matrix in Eq.~(\ref{Lindblad}), we find
\begin{eqnarray}\label{diff3}
\left\{
\begin{array}{l}
\dot{a}(t) = 2k\Big(g(t)+3e(t)-4a(t)\Big),\\
\dot{b}(t) = 4k\Big(c(t)-3b(t)\Big),\\
\dot{c}(t)= 2k\Big(b(t)+d(t)-6c(t)\Big),\\
\dot{d}(t)= 4k\Big(c(t)-3d(t)\Big),\\
\dot{e}(t)= 4k\Big(a(t)+h(t)-2e(t)\Big),\\
\dot{f}(t)= 8k\Big(h(t)-f(t)\Big),\\
\dot{g}(t)= 8k\Big(a(t)-g(t)\Big),\\
\dot{h}(t)= 2k\Big(3e(t)+f(t)-4h(t)\Big),
\end{array}\right.
\end{eqnarray}
in which the initial conditions are $a(0)=b(0)=1/4$ and
$c(0)=d(0)=e(0)=f(0)=g(0)=h(0)=0$. The solutions are found as
\begin{eqnarray}
\label{solution3} \left\{
\begin{array}{l}
a(t) =\frac{1}{16}\left( 1 + e^{-4 \kappa t} + e^{-12 \kappa t} + e^{-16 \kappa t}\right),\\
b(t) =\frac{1}{16}e^{-16 \kappa t}\left( 1+ e^{4 \kappa t}\right)^2,\\
c(t) =\frac{1}{16}e^{-8 \kappa t}\left( 1 -  e^{-8 \kappa t}\right),\\
d(t) =\frac{1}{16}e^{-16 \kappa t}\left( -1+ e^{4 \kappa t}\right)^2,\\
e(t) =\frac{1}{16}\left( 1 - e^{-16 \kappa t}\right),\\
f(t) =\frac{1}{16}\left( 1 - 2e^{-4 \kappa t} + 2e^{-12 \kappa t} - e^{-16 \kappa t}\right),\\
g(t) =\frac{1}{16}\left( 1 + 2e^{-4 \kappa t} - 2e^{-12 \kappa t} - e^{-16 \kappa t}\right),\\
h(t) =\frac{1}{16}\left( 1 - 2e^{-4 \kappa t} - 2e^{-12 \kappa t} + e^{-16 \kappa t}\right).
\end{array}\right.
\end{eqnarray}
The corresponding lower bound for this state is depicted in
Fig.~\ref{fig2} as a black line.

\section{Evolution of entanglement of maximal entangled four-qubit states}\label{sec3}
In this section, we compare the time evolution of entanglement of
$W_4$ state with those of maximally entangled four-qubit states,
namely Eq.~(\ref{1}). For this purpose, we solve the Lindblad
equation (\ref{Lindblad}) for these states  and then compute the
lower bound to the concurrence for each case.

First, let us consider the state $|\phi_1\rangle$ which is indeed
the four-qubit GHZ state.  For this case, the Lindblad equation is
solved for the Pauli channels $\sigma_x$, $\sigma_y$, $\sigma_z$ as
well as the isotropic channel and the lower bound (\ref{lowerbound})
is computed for each case in Ref.~\cite{SP2}. Here, we only present
the results as follows:
\begin{eqnarray}\label{ent}
\tau_x(\phi_1)&=& \tau_y(\phi_1) = \sqrt{2} \,\,\mbox{max} \left\{0, \frac{1}{4} \left(e^{-8 \kappa t} + 6 e^{-4 \kappa t} - 3\right)\right\},\\
\tau_z(\phi_1) &=& \sqrt{2}\,\, e^{-8 \kappa t},\\
\tau_d(\phi_1) &=& \sqrt{2}\,\, \mbox{max} \left\{0, \frac{1}{8}
\left(9 e^{-16 \kappa t} + 6 e^{-8 \kappa t} - 7\right)\right\},
\end{eqnarray}

Now, consider the state $|\phi_2\rangle$ which is transmitted
through the Pauli-X channel. Its time evolution as a solution of the
Lindblad equation (\ref{Lindblad}) and after following the steps
similar to the previous section is obtained as
\begin{eqnarray}\label{fi4x}
\rho_{\phi_2}^x= { \left(
\begin{smallmatrix}
a & 0 & 0 & a & 0 & 0 & 0 & 0 & 0 & 0 & 0 & 0 & 0 & a & a & 0  \\
0 & b & b & 0 & 0 & 0 & 0 & 0 & 0 & 0 & 0 & 0 & b & 0 & 0 & b   \\
0 & b & b & 0 & 0 & 0 & 0 & 0 & 0 & 0 & 0 & 0 & b & 0 & 0 & b   \\
a & 0 & 0 & a & 0 & 0 & 0 & 0 & 0 & 0 & 0 & 0 & 0 & a & a & 0   \\
0 & 0 & 0 & 0 & c & 0 & 0 & c & 0 & c & c & 0 & 0 & 0 & 0 & 0   \\
0 & 0 & 0 & 0 & 0 & c & c & 0 & c & 0 & 0 & c & 0 & 0 & 0 & 0   \\
0 & 0 & 0 & 0 & 0 & c & c & 0 & c & 0 & 0 & c & 0 & 0 & 0 & 0   \\
0 & 0 & 0 & 0 & c & 0 & 0 & c & 0 & c & c & 0 & 0 & 0 & 0 & 0   \\
0 & 0 & 0 & 0 & 0 & c & c & 0 & c & 0 & 0 & c & 0 & 0 & 0 & 0   \\
0 & 0 & 0 & 0 & c & 0 & 0 & c & 0 & c & c & 0 & 0 & 0 & 0 & 0   \\
0 & 0 & 0 & 0 & c & 0 & 0 & c & 0 & c & c & 0 & 0 & 0 & 0 & 0   \\
0 & 0 & 0 & 0 & 0 & c & c & 0 & c & 0 & 0 & c & 0 & 0 & 0 & 0   \\
0 & b & b & 0 & 0 & 0 & 0 & 0 & 0 & 0 & 0 & 0 & b & 0 & 0 & b   \\
a & 0 & 0 & a & 0 & 0 & 0 & 0 & 0 & 0 & 0 & 0 & 0 & a & a & 0   \\
a & 0 & 0 & a & 0 & 0 & 0 & 0 & 0 & 0 & 0 & 0 & 0 & a & a & 0   \\
0 & b & b & 0 & 0 & 0 & 0 & 0 & 0 & 0 & 0 & 0 & b & 0 & 0 & b
\end{smallmatrix} \right),}
\end{eqnarray}
where
\begin{eqnarray} \left\{
\begin{array}{l}
a =\frac{1}{16} \left(1 + e^{-4 \kappa t} -2 e^{-6 \kappa t}\right),\\
b =\frac{1}{16}\left(1 + e^{-4 \kappa t} +2 e^{-6 \kappa t}\right),\\
c=\frac{1}{16} \left(1 - e^{-4 \kappa t}\right).
\end{array} \right.
\end{eqnarray}
For this case, the lower bound (\ref{lowerbound}) becomes
\begin{eqnarray}\label{entfi4x}
\tau_x(\phi_2) = \sqrt{2}\,\, \mbox{max} \left\{0, \frac{1}{2}
\left(2 e^{-6 \kappa t} + e^{-4 \kappa t} - 1\right)\right\}.
\end{eqnarray}

For the next case, the state $|\phi_2\rangle$ is exposed to the
Pauli-Y channel. The solution  reads
\begin{eqnarray}
\rho_{\phi_2}^y= { \left(
\begin{smallmatrix}
a & 0 & 0 & e & 0 & 0 & 0 & 0 & 0 & 0 & 0 & 0 & 0 & n & n & 0  \\
0 & b & g & 0 & 0 & 0 & 0 & 0 & 0 & 0 & 0 & 0 & m & 0 & 0 & m   \\
0 & g & b & 0 & 0 & 0 & 0 & 0 & 0 & 0 & 0 & 0 & m & 0 & 0 & m   \\
e & 0 & 0 & a & 0 & 0 & 0 & 0 & 0 & 0 & 0 & 0 & 0 & n & n & 0   \\
0 & 0 & 0 & 0 & c & 0 & 0 & f & 0 & d & d & 0 & 0 & 0 & 0 & 0   \\
0 & 0 & 0 & 0 & 0 & c & f & 0 & d & 0 & 0 & d & 0 & 0 & 0 & 0   \\
0 & 0 & 0 & 0 & 0 & f & c & 0 & d & 0 & 0 & d & 0 & 0 & 0 & 0   \\
0 & 0 & 0 & 0 & f & 0 & 0 & c & 0 & d & d & 0 & 0 & 0 & 0 & 0   \\
0 & 0 & 0 & 0 & 0 & d & d & 0 & c & 0 & 0 & f & 0 & 0 & 0 & 0   \\
0 & 0 & 0 & 0 & d & 0 & 0 & d & 0 & c & f & 0 & 0 & 0 & 0 & 0   \\
0 & 0 & 0 & 0 & d & 0 & 0 & d & 0 & f & c & 0 & 0 & 0 & 0 & 0   \\
0 & 0 & 0 & 0 & 0 & d & d & 0 & f & 0 & 0 & c & 0 & 0 & 0 & 0   \\
0 & m & m & 0 & 0 & 0 & 0 & 0 & 0 & 0 & 0 & 0 & b & 0 & 0 & g   \\
n & 0 & 0 & n & 0 & 0 & 0 & 0 & 0 & 0 & 0 & 0 & 0 & a & e & 0   \\
n & 0 & 0 & n & 0 & 0 & 0 & 0 & 0 & 0 & 0 & 0 & 0 & e & a & 0   \\
0 & m & m & 0 & 0 & 0 & 0 & 0 & 0 & 0 & 0 & 0 & g & 0 & 0 & b
\end{smallmatrix} \right),}
\end{eqnarray}
so that
\begin{eqnarray} \left\{
\begin{array}{l}
a =\frac{1}{16} \left(1 + e^{-4 \kappa t} -2 e^{-6 \kappa t}\right),\\
b =\frac{1}{16}\left(1 + e^{-4 \kappa t} +2 e^{-6 \kappa t}\right),\\
c=\frac{1}{16} \left(1 - e^{-4 \kappa t}\right),\\
d=\frac{1}{16} \left(e^{-6 \kappa t} - e^{-2 \kappa t}\right),\\
e =-\frac{1}{16} \left(2 e^{-2 \kappa t} - e^{-4 \kappa t} - e^{-8 \kappa t}\right),\\
f=\frac{1}{16} \left(e^{-4 \kappa t} - e^{-8 \kappa t}\right),\\
g =\frac{1}{16} \left(2 e^{-2 \kappa t} + e^{-4 \kappa t} + e^{-8 \kappa t}\right),\\
m =\frac{1}{16}\left(e^{-2 \kappa t} + 2 e^{-4 \kappa t} + e^{-6 \kappa t}\right),\\
n =\frac{1}{16}\left(e^{-2 \kappa t} - 2 e^{-4 \kappa t} + e^{-6 \kappa t}\right).
\end{array} \right.
\end{eqnarray}
The lower bound (\ref{lowerbound}) now becomes
\begin{eqnarray}\label{entfi4y}
\tau_y(\phi_2) = \sqrt{2}\,\, \mbox{max} \left\{0, \frac{1}{8}
\left(e^{-8 \kappa t} + 4 e^{-6 \kappa t} +6 e^{-4 \kappa t} + 4
e^{-2 \kappa t} - 7\right)\right\}.
\end{eqnarray}

For transmission of $|\phi_2\rangle$ through the Pauli-Z channel,
the time evolution of the corresponding density matrix is
\begin{eqnarray}
\rho_{\phi_2}^z= { \left(
\begin{smallmatrix}
0 & 0 & 0 & 0 & 0 & 0 & 0 & 0 & 0 & 0 & 0 & 0 & 0 & 0 & 0 & 0  \\
0 & a & c & 0 & 0 & 0 & 0 & 0 & 0 & 0 & 0 & 0 & b & 0 & 0 & b   \\
0 & c & a & 0 & 0 & 0 & 0 & 0 & 0 & 0 & 0 & 0 & b & 0 & 0 & b   \\
0 & 0 & 0 & 0 & 0 & 0 & 0 & 0 & 0 & 0 & 0 & 0 & 0 & 0 & 0 & 0   \\
0 & 0 & 0 & 0 & 0 & 0 & 0 & 0 & 0 & 0 & 0 & 0 & 0 & 0 & 0 & 0   \\
0 & 0 & 0 & 0 & 0 & 0 & 0 & 0 & 0 & 0 & 0 & 0 & 0 & 0 & 0 & 0   \\
0 & 0 & 0 & 0 & 0 & 0 & 0 & 0 & 0 & 0 & 0 & 0 & 0 & 0 & 0 & 0   \\
0 & 0 & 0 & 0 & 0 & 0 & 0 & 0 & 0 & 0 & 0 & 0 & 0 & 0 & 0 & 0   \\
0 & 0 & 0 & 0 & 0 & 0 & 0 & 0 & 0 & 0 & 0 & 0 & 0 & 0 & 0 & 0   \\
0 & 0 & 0 & 0 & 0 & 0 & 0 & 0 & 0 & 0 & 0 & 0 & 0 & 0 & 0 & 0   \\
0 & 0 & 0 & 0 & 0 & 0 & 0 & 0 & 0 & 0 & 0 & 0 & 0 & 0 & 0 & 0   \\
0 & 0 & 0 & 0 & 0 & 0 & 0 & 0 & 0 & 0 & 0 & 0 & 0 & 0 & 0 & 0   \\
0 & b & b & 0 & 0 & 0 & 0 & 0 & 0 & 0 & 0 & 0 & a & 0 & 0 & c   \\
0 & 0 & 0 & 0 & 0 & 0 & 0 & 0 & 0 & 0 & 0 & 0 & 0 & 0 & 0 & 0   \\
0 & 0 & 0 & 0 & 0 & 0 & 0 & 0 & 0 & 0 & 0 & 0 & 0 & 0 & 0 & 0   \\
0 & b & b & 0 & 0 & 0 & 0 & 0 & 0 & 0 & 0 & 0 & c & 0 & 0 & a
\end{smallmatrix} \right),}
\end{eqnarray}
where
\begin{eqnarray} \left\{
\begin{array}{l}
a =\frac{1}{4},\\
b =\frac{1}{4} e^{-6 \kappa t},\\
c=\frac{1}{4} e^{-4 \kappa t}.
\end{array} \right.
\end{eqnarray}
The lower bound to the concurrence for this case is similar to the
lower bound obtained for Pauli-X channel, namely $\tau_z(\phi_2)
=\tau_x(\phi_2)$.

Finally, if $|\phi_2\rangle$ is exposed to the depolarizing channel,
its density matrix takes the following form
\begin{eqnarray}
\rho_{\phi_2}^d= { \left(
\begin{smallmatrix}
a & 0 & 0 & e & 0 & 0 & 0 & 0 & 0 & 0 & 0 & 0 & 0 & n & n & 0  \\
0 & b & d & 0 & 0 & 0 & 0 & 0 & 0 & 0 & 0 & 0 & m & 0 & 0 & m   \\
0 & d & b & 0 & 0 & 0 & 0 & 0 & 0 & 0 & 0 & 0 & m & 0 & 0 & m   \\
e & 0 & 0 & a & 0 & 0 & 0 & 0 & 0 & 0 & 0 & 0 & 0 & n & n & 0   \\
0 & 0 & 0 & 0 & c & 0 & 0 & f & 0 & 0 & 0 & 0 & 0 & 0 & 0 & 0   \\
0 & 0 & 0 & 0 & 0 & c & f & 0 & 0 & 0 & 0 & 0 & 0 & 0 & 0 & 0   \\
0 & 0 & 0 & 0 & 0 & f & c & 0 & 0 & 0 & 0 & 0 & 0 & 0 & 0 & 0   \\
0 & 0 & 0 & 0 & f & 0 & 0 & c & 0 & 0 & 0 & 0 & 0 & 0 & 0 & 0   \\
0 & 0 & 0 & 0 & 0 & 0 & 0 & 0 & c & 0 & 0 & f & 0 & 0 & 0 & 0   \\
0 & 0 & 0 & 0 & 0 & 0 & 0 & 0 & 0 & c & f & 0 & 0 & 0 & 0 & 0   \\
0 & 0 & 0 & 0 & 0 & 0 & 0 & 0 & 0 & f & c & 0 & 0 & 0 & 0 & 0   \\
0 & 0 & 0 & 0 & 0 & 0 & 0 & 0 & f & 0 & 0 & c & 0 & 0 & 0 & 0   \\
0 & m & m & 0 & 0 & 0 & 0 & 0 & 0 & 0 & 0 & 0 & b & 0 & 0 & d   \\
n & 0 & 0 & n & 0 & 0 & 0 & 0 & 0 & 0 & 0 & 0 & 0 & a & e & 0   \\
n & 0 & 0 & n & 0 & 0 & 0 & 0 & 0 & 0 & 0 & 0 & 0 & e & a & 0   \\
0 & m & m & 0 & 0 & 0 & 0 & 0 & 0 & 0 & 0 & 0 & d & 0 & 0 & b
\end{smallmatrix} \right),}
\end{eqnarray}
where
\begin{eqnarray} \left\{
\begin{array}{l}
a =\frac{1}{16} \left(1 + e^{-8 \kappa t} -2 e^{-12 \kappa t}\right),\\
b =\frac{1}{16}\left(1 + e^{-8 \kappa t} +2 e^{-12 \kappa t}\right),\\
c=\frac{1}{16} \left(1 - e^{-8 \kappa t}\right),\\
d =\frac{1}{16} \left(e^{-8 \kappa t} + 2 e^{-12 \kappa t} + e^{-16 \kappa t}\right), \\
e =-\frac{1}{16} \left(e^{-8 \kappa t} - 2 e^{-12 \kappa t} + e^{-16 \kappa t}\right), \\
f=\frac{1}{16} \left(e^{-8 \kappa t} - e^{-16 \kappa t}\right),\\
m =\frac{1}{8}\left(e^{-12 \kappa t} +  e^{-16 \kappa t}\right),\\
n =\frac{1}{8}\left(e^{-12 \kappa t} -  e^{-16 \kappa t}\right).
\end{array} \right.
\end{eqnarray}
In this case, Eq.~(\ref{lowerbound}) gives
\begin{eqnarray}\label{entfi4d}
\tau_d(\phi_2) = \sqrt{2}\,\, \mbox{max} \left\{0, \frac{1}{8}
\left(7 e^{-16 \kappa t} + 8 e^{-12 \kappa t} - 7\right)\right\}.
\end{eqnarray}

\begin{figure}[htp]
  \centering
  \begin{tabular}{cc}
{\includegraphics[width=80mm]{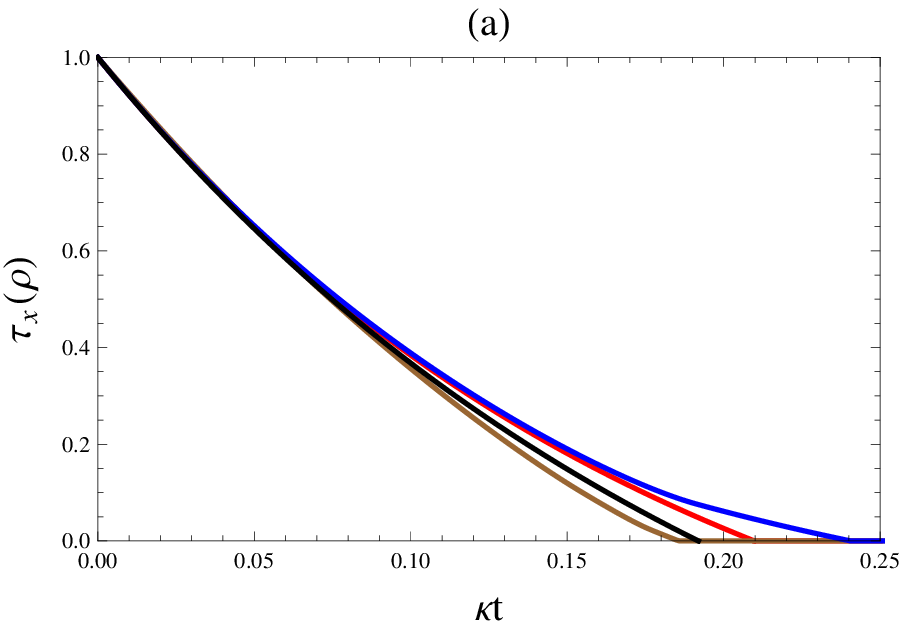}}&
{\includegraphics[width=80mm]{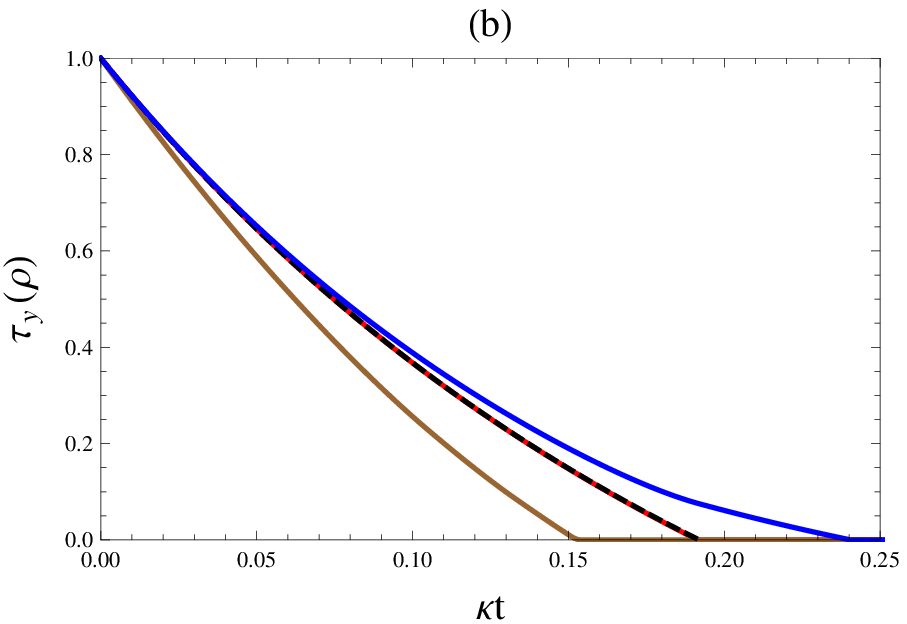}}\\
{\includegraphics[width=80mm]{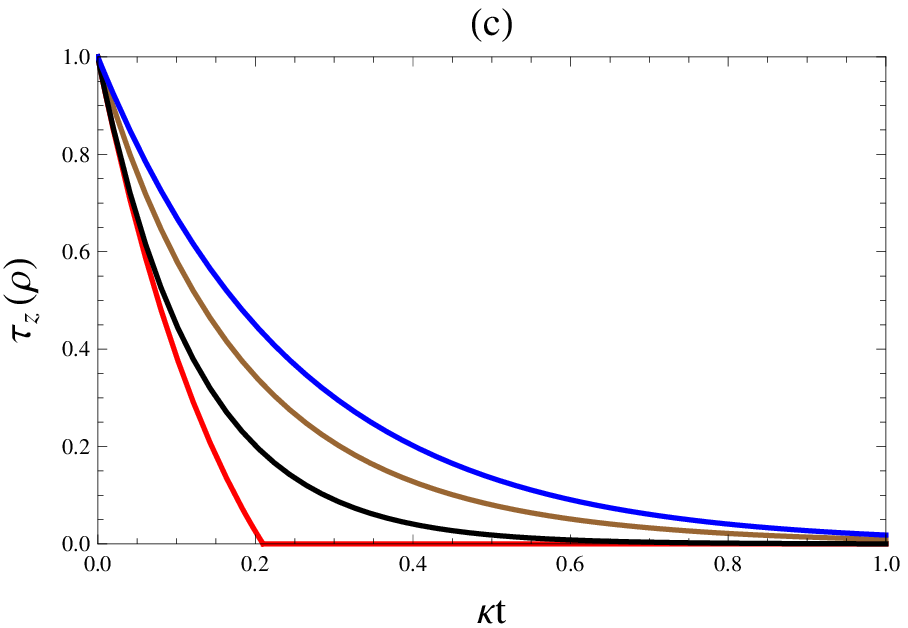}}&
{\includegraphics[width=80mm]{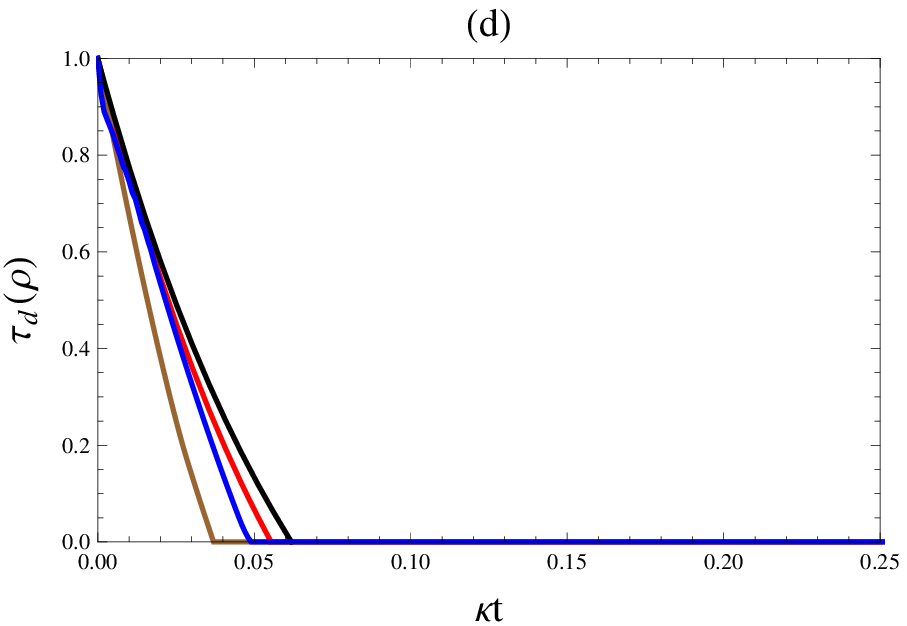}}
\end{tabular}
     \caption{\label{fig3} The lower bound to the concurrence
for $W_4$ (blue line), four-qubit GHZ (black line), $\phi_2$ (red
line) and $\phi_3$ (brown line)  under: (a) Pauli-X (b) Pauli-Y (c)
Pauli-Z and (d) isotropic noisy channels versus $\kappa t$.}
\end{figure}

Now consider the state $|\phi_3\rangle$. If this state is
transmitted through Pauli-X channel, the time evolution of its
density matrix is obtained as
\begin{eqnarray}
\rho_{\phi_3}^x= { \left(
\begin{smallmatrix}
a & 0 & 0 & d & 0 & d & d & q & 0 & d & d & q & d & q & q & 0 \\
0 & g & b & 0 & b & 0 & n & m & b & 0 & n & m & n & m & 0 & c \\
0 & b & g & 0 & b & n & 0 & m & b & n & 0 & m & n & 0 & m & c \\
d & 0 & 0 & p & n & f & f & 0 & n & f & f & 0 & 0 & e & e & u \\
0 & b & b & n & g & 0 & 0 & m & b & n & n & 0 & 0 & m & m & c \\
d & 0 & n & f & 0 & p & f & 0 & n & f & 0 & e & f & 0 & e & u \\
d & n & 0 & f & 0 & f & p & 0 & n & 0 & f & e & f & e & 0 & u \\
q & m & m & 0 & m & 0 & 0 & l & 0 & e & e & r & e & r & r & 0 \\
0 & b & b & n & b & n & n & 0 & g & 0 & 0 & m & 0 & m & m & c \\
d & 0 & n & f & n & f & 0 & e & 0 & p & f & 0 & f & 0 & e & u \\
d & n & 0 & f & n & 0 & f & e & 0 & f & p & 0 & f & e & 0 & u \\
q & m & m & 0 & 0 & e & e & r & m & 0 & 0 & l & e & r & r & 0 \\
d & n & n & 0 & 0 & f & f & e & 0 & f & f & e & p & 0 & 0 & u \\
q & m & 0 & e & m & 0 & e & r & m & 0 & e & r & 0 & l & r & 0 \\
q & 0 & m & e & m & e & 0 & r & m & e & 0 & r & 0 & r & l & 0 \\
0 & c & c & u & c & u & u & 0 & c & u & u & 0 & u & 0 & 0 & h
\end{smallmatrix} \right),}
\end{eqnarray}
so that
\begin{eqnarray}\label{x}
\left\{
\begin{array}{l}
a =\frac{1}{48} \left(3 +6 e^{-4 \kappa t} -8 e^{-6 \kappa t} -e^{-8 \kappa t}\right), \\
b =\frac{1}{48}\left(1 +2 e^{-2 \kappa t}+2 e^{-4 \kappa t} +2 e^{-6 \kappa t} + e^{-8 \kappa t}\right),\\
d=\frac{1}{48} \left(1 +2 e^{-2 \kappa t}-2 e^{-6 \kappa t} - e^{-8 \kappa t}\right), \\
f=\frac{1}{48} \left(1 - e^{-8 \kappa t}\right),\\
g =\frac{1}{48} \left(3 +4 e^{-6 \kappa t} + e^{-8 \kappa t}\right),\\
h =\frac{1}{48}\left(3 +6 e^{-4 \kappa t} +8 e^{-6 \kappa t} -e^{-8 \kappa t}\right), \\
l =\frac{1}{48} \left(3 -4 e^{-6 \kappa t} + e^{-8 \kappa t}\right), \\
m =\frac{1}{48} \left(1 -2 e^{-4 \kappa t} + e^{-8 \kappa t}\right), \\
n =-\frac{1}{24\sqrt{2}} \left(1 - e^{-2 \kappa t} - e^{-4 \kappa t}
+ e^{-6 \kappa t}\right).\\
c=\frac{1}{24\sqrt{2}} \left(1 + e^{-2 \kappa t}+3 e^{-4 \kappa t} +3 e^{-6 \kappa t}\right),\\
e =-\frac{1}{24\sqrt{2}} \left(1 + e^{-2 \kappa t} - e^{-4 \kappa t} - e^{-6 \kappa t}\right),\\
q =\frac{1}{24\sqrt{2}} \left(1 - e^{-2 \kappa t}+3 e^{-4 \kappa t}
-3 e^{-6 \kappa t}\right).
\end{array} \right.
\end{eqnarray}
The lower bound for this density matrix is depicted in
Fig.~\ref{fig3}(a).

The solution of the Lindblad equation when $|\phi_3\rangle$ is
exposed Pauli-Y noise is given by
\begin{eqnarray}
\rho_{\phi_3}^y= { \left(
\begin{smallmatrix}
a & 0 & 0 & -d & 0 & -d & -d & q' & 0 & -d & -d & q' & -d & q' & q' & 0 \\
0 & g & b & 0 & b & 0 & n & -m & b & 0 & n & -m & n & -m & 0 & c' \\
0 & b & g & 0 & b & n & 0 & -m & b & n & 0 & -m & n & 0 & -m & c' \\
-d & 0 & 0 & p & n & f & f & 0 & n & f & f & 0 & 0 & e' & e' & u \\
0 & b & b & n & g & 0 & 0 & -m & b & n & n & 0 & 0 & -m & -m & c' \\
-d & 0 & n & f & 0 & p & f & 0 & n & f & 0 & e' & f & 0 & e' & -u \\
-d & n & 0 & f & 0 & f & p & 0 & n & 0 & f & e' & f & e' & 0 & -u \\
q' & -m & -m & 0 & -m & 0 & 0 & l & 0 & e' & e' & r & e' & r & r & 0 \\
0 & b & b & n & b & n & n & 0 & g & 0 & 0 & -m & 0 & -m & -m & c' \\
-d & 0 & n & f & n & f & 0 & e' & 0 & p & f & 0 & f & 0 & e' & -u \\
-d & n & 0 & f & n & 0 & f & e' & 0 & f & p & 0 & f & e' & 0 & -u \\
q' & -m & -m & 0 & 0 & e' & e' & r & -m & 0 & 0 & l & e' & r & r & 0 \\
-d & n & n & 0 & 0 & f & f & e' & 0 & f & f & e' & p & 0 & 0 & -u \\
q' & -m & 0 & e' & -m & 0 & e' & r & -m & 0 & e' & r & 0 & l & r & 0 \\
q' & 0 & -m & e' & -m & e' & 0 & r &- m & e' & 0 & r & 0 & r & l & 0 \\
0 & c' & c' & u & c' & -u & -u & 0 & c' & -u & -u & 0 & -u & 0 & 0 & h
\end{smallmatrix} \right),}
\end{eqnarray}
where $a$, $b$, $d$, $f$, $g$, $h$, $l$, $m$, $n$, $p$, $r$ and $u$
are given in Eq.~(\ref{x}) and $c'$, $e'$ and $q'$ are found as
\begin{eqnarray} \left\{
\begin{array}{l}
c'=\frac{1}{24\sqrt{2}} \left(3 e^{-2 \kappa t}+3 e^{-4 \kappa t} + e^{-6 \kappa t} + e^{-8 \kappa t}\right),\\
e' =-\frac{1}{24\sqrt{2}} \left(e^{-2 \kappa t} + e^{-4 \kappa t} - e^{-6 \kappa t} - e^{-8 \kappa t}\right), \\
q' =\frac{1}{24\sqrt{2}} \left(3 e^{-2 \kappa t} -3 e^{-4 \kappa t} + e^{-6 \kappa t} - e^{-8 \kappa t}\right).
\end{array} \right.
\end{eqnarray}
The lower bound for this density matrix is shown in
Fig.~\ref{fig3}(b).

Now, let us consider the state $|\phi_3\rangle$ which is transmitted
through Pauli-Z channel. After solving the Lindblad equation, the
time evolution of the density matrix is obtained as
\begin{eqnarray}
\rho_{\phi_3}^z= { \left(
\begin{smallmatrix}
0 & 0 & 0 & 0 & 0 & 0 & 0 & 0 & 0 & 0 & 0 & 0 & 0 & 0 & 0 & 0  \\
0 & a & b & 0 & b & 0 & 0 & 0 & b & 0 & 0 & 0 & 0 & 0 & 0 & c   \\
0 & b & a & 0 & b & 0 & 0 & 0 & b & 0 & 0 & 0 & 0 & 0 & 0 & c   \\
0 & 0 & 0 & 0 & 0 & 0 & 0 & 0 & 0 & 0 & 0 & 0 & 0 & 0 & 0 & 0   \\
0 & b & b & 0 & a & 0 & 0 & 0 & b & 0 & 0 & 0 & 0 & 0 & 0 & c   \\
0 & 0 & 0 & 0 & 0 & 0 & 0 & 0 & 0 & 0 & 0 & 0 & 0 & 0 & 0 & 0   \\
0 & 0 & 0 & 0 & 0 & 0 & 0 & 0 & 0 & 0 & 0 & 0 & 0 & 0 & 0 & 0   \\
0 & 0 & 0 & 0 & 0 & 0 & 0 & 0 & 0 & 0 & 0 & 0 & 0 & 0 & 0 & 0   \\
0 & b & b & 0 & b & 0 & 0 & 0 & a & 0 & 0 & 0 & 0 & 0 & 0 & c   \\
0 & 0 & 0 & 0 & 0 & 0 & 0 & 0 & 0 & 0 & 0 & 0 & 0 & 0 & 0 & 0   \\
0 & 0 & 0 & 0 & 0 & 0 & 0 & 0 & 0 & 0 & 0 & 0 & 0 & 0 & 0 & 0   \\
0 & 0 & 0 & 0 & 0 & 0 & 0 & 0 & 0 & 0 & 0 & 0 & 0 & 0 & 0 & 0   \\
0 & 0 & 0 & 0 & 0 & 0 & 0 & 0 & 0 & 0 & 0 & 0 & 0 & 0 & 0 & 0   \\
0 & 0 & 0 & 0 & 0 & 0 & 0 & 0 & 0 & 0 & 0 & 0 & 0 & 0 & 0 & 0   \\
0 & 0 & 0 & 0 & 0 & 0 & 0 & 0 & 0 & 0 & 0 & 0 & 0 & 0 & 0 & 0   \\
0 & c & c & 0 & c & 0 & 0 & 0 & c & 0 & 0 & 0 & 0 & 0 & 0 & d
\end{smallmatrix} \right),}
\end{eqnarray}
where
\begin{eqnarray} \left\{
\begin{array}{l}
a =\frac{1}{6}, \\
b =\frac{1}{6} e^{-4 \kappa t},\\
c=\frac{1}{3\sqrt{2}} e^{-6 \kappa t}, \\
d =\frac{1}{3}.
\end{array} \right.
\end{eqnarray}
Computation of the lower bound gives rise to
\begin{eqnarray}\label{entfi5z}
\tau_z(\phi_3) =\frac{2}{3} \sqrt{\left(3 e^{-12 \kappa t} + e^{-8 \kappa t}\right)}.
\end{eqnarray}

To this end, consider the transmission of $|\phi_3\rangle$ through
the depolarizing channel. In this case, the time evolution of the
state is described by
\begin{eqnarray}
\rho_{\phi_3}^d= { \left(
\begin{smallmatrix}
a & 0 & 0 & 0 & 0 & 0 & 0 & e & 0 & 0 & 0 & e & 0 & e & e & 0  \\
0 & d & b & 0 & b & 0 & 0 & 0 & b & 0 & 0 & 0 & 0 & 0 & 0 & c   \\
0 & b & d & 0 & b & 0 & 0 & 0 & b & 0 & 0 & 0 & 0 & 0 & 0 & c   \\
0 & 0 & 0 & g & 0 & m & m & 0 & 0 & m & m & 0 & 0 & 0 & 0 & 0   \\
0 & b & b & 0 & d & 0 & 0 & 0 & b & 0 & 0 & 0 & 0 & 0 & 0 & c   \\
0 & 0 & 0 & m & 0 & g & m & 0 & 0 & m & 0 & 0 & m & 0 & 0 & 0   \\
0 & 0 & 0 & m & 0 & m & g & 0 & 0 & 0 & m & 0 & m & 0 & 0 & 0   \\
e & 0 & 0 & 0 & 0 & 0 & 0 & f & 0 & 0 & 0 & n & 0 & n & n & 0   \\
0 & b & b & 0 & b & 0 & 0 & 0 & d & 0 & 0 & 0 & 0 & 0 & 0 & c   \\
0 & 0 & 0 & m & 0 & m & 0 & 0 & 0 & g & m & 0 & m & 0 & 0 & 0   \\
0 & 0 & 0 & m & 0 & 0 & m & 0 & 0 & m & g & 0 & m & 0 & 0 & 0   \\
e & 0 & 0 & 0 & 0 & 0 & 0 & n & 0 & 0 & 0 & f & 0 & n & n & 0   \\
0 & 0 & 0 & 0 & 0 & m & m & 0 & 0 & m & m & 0 & g & 0 & 0 & 0   \\
e & 0 & 0 & 0 & 0 & 0 & 0 & n & 0 & 0 & 0 & n & 0 & f & n & 0   \\
e & 0 & 0 & 0 & 0 & 0 & 0 & n & 0 & 0 & 0 & n & 0 & n & f & 0   \\
0 & c & c & 0 & c & 0 & 0 & 0 & c & 0 & 0 & 0 & 0 & 0 & 0 & h
\end{smallmatrix} \right),}
\end{eqnarray}
so that
\begin{eqnarray}
\left\{
\begin{array}{l}
a=\frac{1}{48} \left(3 +6 e^{-8 \kappa t} -8 e^{-12 \kappa t} -e^{-16 \kappa t}\right),\\
b =\frac{1}{24}\left(e^{-8 \kappa t} +2 e^{-12 \kappa t} + e^{-16 \kappa t}\right),\\
c=\frac{1}{6\sqrt{2}} \left(e^{-12 \kappa t} + e^{-16 \kappa t}\right),\\
d=\frac{1}{48} \left(3 +4 e^{-12 \kappa t} + e^{-16 \kappa t}\right), \\
e=\frac{1}{6\sqrt{2}} \left(e^{-12 \kappa t} - e^{-16 \kappa t}\right),\\
f=\frac{1}{48} \left(3 -4 e^{-12 \kappa t} + e^{-16 \kappa t}\right),\\
h =\frac{1}{48}\left(3 +6 e^{-8 \kappa t} +8 e^{-12 \kappa t} -e^{-16 \kappa t}\right),\\
m =\frac{1}{24} \left(e^{-8 \kappa t} - e^{-16 \kappa t}\right),\\
n =\frac{1}{24}\left(e^{-8 \kappa t} -2 e^{-12 \kappa t} + e^{-16 \kappa t}\right).
\end{array}\right.
\end{eqnarray}
The lower bound to the concurrence is plotted in Fig.~\ref{fig3}(d).

In Fig.~(\ref{fig3}), we plotted the results for the four
investigated states. It should be noted that although the  states
introduced in Eq.~(\ref{1}) are maximally entangled, the initial
lower bound to the concurrence for these states gives rise to
different values, i.e., $\tau(\phi_1)=\tau(\phi_2)=\sqrt{2}$ and
$\tau(\phi_3)=4/3$. So, we normalized them so that
$\tau(\phi_1)=\tau(\phi_2)=\tau(\phi_3)=1$ at $\kappa t=0$. As it is
shown in Fig.~(\ref{fig3}),  the entanglement present in $W_4$ state
is more robust than the other investigated states against
decoherence under bit-flip, bit-phase-flip, and phase-flip noises.
However, the four-qubit GHZ state is more robust in the presence of
the depolarizing noise.

\section{Conclusions}\label{sec4}
In this paper, we have investigated the dynamics of an initial
four-qubit W state in interaction with its surrounding environment.
We exactly solved the Lindblad equation in which the Lindblad
operators are proportional to the Pauli matrices and obtained the
density matrices corresponding to each noisy channel. We also
examined the time evolution of entanglement using the lower bound to
the concurrence for four-qubit W state. It is found that the
entanglement of states vanishes after some finite time for the Pauli
channels $\sigma_x$ and $\sigma_y$ as well as the isotropic channel.
However, for the Pauli-Z noise, the entanglement exponentially
decreases and vanishes asymptotically. Moreover, by exactly solving
the Lindblad equation, we studied the effects of various noises on
three four-qubit maximally  entangled states and obtained the time
evolution of the lower bound. We found that except the depolarizing
channel, the $W_4$ state is more robust against the decoherence with
respect to $\phi_1$, $\phi_2$, and $\phi_3$. Also, except Pauli-Z
channel, $W_3$  is more robust than $W_4$ under the noises
\cite{siomau10}.

\end{document}